\documentclass[conference]{IEEEtran}
\IEEEoverridecommandlockouts

\usepackage{amssymb}
\usepackage[cmex10]{amsmath}
\usepackage{stfloats}
\usepackage{graphicx}
\usepackage{subfig}
\usepackage{tabularx}
\usepackage{epsfig,epsf,color,balance,cite}
\usepackage{verbatim}
\usepackage{url}
\usepackage{bm}
\usepackage{booktabs}

\usepackage{algorithm}
\usepackage{algorithmic}

\usepackage{color}

\definecolor{myc1}{rgb}{0,0,0}
\allowdisplaybreaks[2]
\begin{document}
\title{Rate Maximization for  Fluid Antenna System Assisted Semantic Communication}

\author{Siyun Liang,  
            Chen Zhu,
            Zhaohui Yang,
            Changsheng You,
            Dusit Niyato, \IEEEmembership{Fellow, IEEE,}\\
            Kai-Kit Wong, \IEEEmembership{Fellow, IEEE,}
           and Zhaoyang Zhang, \IEEEmembership{Senior Member, IEEE}
           
\vspace{-1em}
\thanks{S. Liang, C. Zhu is with Polytechnic Institute, Zhejiang University, Hangzhou, Zhejiang,310015, China (e-mail: {siyunliang, zhuc}@zju.edu.cn).}
\thanks{Z. Yang, and Z. Zhang are with the College of Information Science and Electronic Engineering, Zhejiang University, and also with Zhejiang Provincial Key Laboratory of Info. Proc., Commun. \& Netw. (IPCAN), Hangzhou, 310027, China (e-mails: \{ yang\_zhaohui, zhzy\}@zju.edu.cn).}
\thanks{C. You is with Department of Electronic and Electrical Engineering Southern University of Science and Technology (SUSTech) Shenzhen, 518055, China (e-mail: youcs@sustech.edu.cn).}
\thanks{Dusit Niyato is with School of Computer Science and Engineering, Nanyang Technological University, 639798, Singapore (e-mail: dniyato@ntu.edu.sg).}
\thanks{Kai-Kit Wong is with the Department of Electronic and Electrical Engineering, University College London, WC1E 7JE London, U.K., and also with the Yonsei Frontier Laboratory, Yonsei University, Seoul 03722, South Korea (e-mail: kai-kit.wong@ucl.ac.uk).}
}

\maketitle
\begin{abstract}
    In this paper, we investigate the problem of rate maximization in a fluid antenna system (FAS) assisted 
 semantic communication system. In the considered model, a base station (BS) with multiple static antennas employs semantic extraction techniques to compress the data ready to be sent to a user.  The user equipped with a fluid antenna is located in the near field coverage region of the BS. Our aim is to jointly optimize the transmit beamforming and the semantic compression rate at the BS, as well as the selection of activated ports in FAS, to maximize the equivalent transmission ratio under a specific power budget. We design an alternating algorithm to solve the problem, where we obtain the optimal semantic compression ratio is in closed form at each step. Simulation results validate the effectiveness of the proposed algorithm.
\end{abstract}
\vspace{-0.5em}
\section{Introduction}

Fluid antenna system (FAS) is proposed as an effective approach to meet the growing demands of novel wireless communication systems. To improve the communication performance, FAS has the ability of reconfiguring the antennas' flexible dimensions or shapes and can be controlled by software. To demonstrate the superiority of FAS, Wong et.al analyzed the potential of FAS surpassing other antenna systems in \cite{9131873}. In practice, there are two main trends of implementing FAS, i. e., liquid-based and pixel-based \cite{9982508}. {   So far, various studies have been conducted on FAS to further investigate the transmission secrecy\cite{10092780}, beamforming and port selection in near-field scenario\cite{Chen2024JointBA}, and FAS for mmWave networks \cite{10167904}.}

{  New communication paradigms are aroused to challenge Shannon's limitation, such as novel modulating paradigm\cite{9963681} and semantic communication (SC). Instead of bit-level correctness, SC aims at guaranteeing the meaning of the information being conveyed correctly. There has been various artificial intelligence (AI) driven implement for SC for text, speech and multi-media \cite{10614204}. To further express the semantic information hidden, knowledge graph and probability graph are introduced \cite{10333452}, representing semantic information by including different entities and their corresponding relations. With abundant techniques to use, the resource allocation in SC is aroused as a crucial challenge\cite{10915662}.  However, there lacks the investigation of FAS assisted SC, even though FAS can further improve the transmission rate.}

 The objective of this paper is to jointly design the beamforming at a base station (BS) and the selection of activated ports at the user. The main contributions are as follows.
\begin{itemize}
    \item We consider a downlink system where a multi-antenna BS implementing semantic information extraction techniques, communicates with a user equipped with an  FAS. The FAS has non-unique activated ports with varying positions. We also formulate an optimization problem to maximize the equivalent rate between the BS and the user, with specific power budget.
    \item To address the rate maximization problem, we take the equivalent rate upper bound to simplify the problem, and decompose it into two sub-problems. We design and propose an algorithm which solves the two sub-problems alternately to obtain a final solution: the first sub-problem concerns optimizing the semantic compression rate and the covariance matrix. This problem is fractional and non-convex, and we solve the problem by using Dinkelbach’s transform. For the second sub-problem aiming at optimizing port selection, it is an integer optimization problem and we introduce an enumerate-like alternative optimization algorithm, which keeps adjusting one single component of the vector variable in its value range.
    \item To prove the effectiveness of the proposed algorithm, numerical results show that the proposed algorithm can achieve higher rate than conventional system without fluid antenna.
\end{itemize}
\vspace{-1em}
\section{System Model}
Consider a downlink communication system, in which the BS and the user have access to shared probability graphs. The BS is constructed with $N$ fixed-position antennas and the user is equipped with an FAS with $M$ ports, which can be activated at any given moment. { The transmit data will be send after semantic extraction. We consider the presence of scatterers in the channel and choose mmWave as the carrier, providing high-speed communication for mobile user. Thus, near-field scenario will be adopted during transmission modeling. }
\subsection{Antenna Coordinate System}\label{sec:A}
{  A coordinate system is established with the center of the transmit antenna array as the origin $O_{t}$. The transmit antennas are distributed uniformly perpendicular to the horizontal plane (denoted by the $x$-axis). The distance between adjacent antennas is $d_{BS}$. The $n$-th antenna's coordinate can be expressed as $(0,y_{{\rm BS},n})$, where $y_{{\rm BS},n}=\frac{2(n-1)-N+1}{2} d_{BS}$.}

{  Similarly, we also set a coordinate system with the center of the user's FAS as the origin $O_r$. During transmission, the FAS activate $m_a$ ports for receiving, with $1\leq m_a\leq M$.  The selected activated ports are represented as $\boldsymbol{r}=[r_1, \cdots, r_m, \cdots, r_{m_a}]^T \in \mathbb{Z}^{m_a\times 1}$, where $r_m \in \{1,2,\ldots,M\},\;r_1<r_2<\cdots<r_{m_a}$. The coordinate of the $r_m$-th port is $(0,y_{{\rm U},m})$, in which $y_{{\rm U},m}=\frac{2(r_m-1)-M+1}{2} d_{U}$.}

\subsection{Signal Transmission Model}\label{sec:C}
The transmit signal is denoted by $\boldsymbol{x} \in \mathbb{C}^{N\times1}\sim\mathcal{C}\mathcal{N}(\boldsymbol{0},\boldsymbol{Q})$ where $\boldsymbol{Q}\in\mathbb{C}^{N\times N}$ represents the transmit covariance matrix. Then, the received signal is expressed as:
{\setlength\abovedisplayskip{3pt}
\setlength\belowdisplayskip{3pt}\begin{equation}
\boldsymbol{y}(\boldsymbol{r})=\boldsymbol{G}(\boldsymbol{r})\boldsymbol{x}+\boldsymbol{w},
\end{equation}}
where $\boldsymbol{G}(\boldsymbol{r})\in \mathbb{C}^{m_a\times N}$ is the channel coefficient matrix from the transmit antennas to the activated ports in the FAS, and $\boldsymbol{w}\in\mathbb{C}^{m_a\times 1}\sim \mathcal{C}\mathcal{N}(\boldsymbol{0},\sigma^2\boldsymbol{\mathrm{I}}_{m_a})$ is the complex additive white Gaussian noise with variance $\sigma^2$ for each element in $\boldsymbol w$.

The number of transmit and receive paths are respectively denoted by $V_t$ and $V_r$. For the $p$-th transmit path, the elevation and azimuth angles of departure (AoDs) are respectively denoted by $\theta_t^p \in [-\frac{\pi}{2},\frac{\pi}{2}]$ and $\phi_t^p \in [0,2\pi]$, while the distance from the scatterer to $O_t$ is denoted by $v_t^p$, and the propagation path difference between the position of the $n$-th transmit antenna and the origin $O_t$ is
{\setlength\abovedisplayskip{3pt}
\setlength\belowdisplayskip{3pt}\begin{equation}\label{eq:dt}
d_t^p(n)=\sqrt{{v_t^p}^2+(y_{{\rm BS},n})^2-2v_t^p y_{{\rm BS},n}\cos\left({\frac{\pi}{2}-\theta_t^p}\right)}-v_t^p.
\end{equation}}

To simplify \eqref{eq:dt}, we use the Talor's approximation $\sqrt{1+x}\approx 1+\frac{1}{2}x-\frac{1}{8}x^2$ for small $x$, and hence we obtain $d_t^p(n)$ as follows:
\vspace{-2mm}
{\setlength\abovedisplayskip{3pt}
\setlength\belowdisplayskip{3pt}\begin{equation}\label{eq:dtn}
d_t^p(n)=-y_{{\rm BS},n} \sin{\theta_t^p}-\frac{y^2_{{\rm BS},n}\sin^2{\theta_t^p}}{2v_t^p}.
\end{equation}}

By employing the ratio of the propagation path difference $d_t^p(n)$ to the carrier wavelength $\lambda$, it is possible to derive the signal phase difference between the position of the $n$-th transmit antenna and origin $O_t$ in the $p$-th transmit path as $2\pi d_t^p(n)/\lambda$. Hence, the transmit field response vector can be written as:
\vspace{-1mm}
\begin{equation}
\boldsymbol{a}(n)\triangleq\left[e^{j\frac{2\pi}{\lambda}d_t^1(n)}, \ldots, e^{j\frac{2\pi}{\lambda}d_t^{V_t}(n)}\right]^T\in \mathbb{C}^{V_t\times1}.
\vspace{-1mm}
\end{equation}

Furthermore, the field response of all the $N$ transmit antennas can be obtained in matrix form as follows:
\begin{equation}
\boldsymbol{A}\triangleq[\boldsymbol{a}(1), \boldsymbol{a}(2), \ldots, \boldsymbol{a}(N)]\in\mathbb{C}^{V_t\times N}.
\end{equation}

We then obtain the receiving field response matrix. In the $q$-th receive path, the elevation and azimuth angles of arrival (AoAs) are respectively denoted by $\theta_r^q \in [-\frac{\pi}{2},\frac{\pi}{2}]$ and $\phi_r^q \in [0,2\pi]$, the distance from the scatterers to the origin $O_r$ in the $q$-th path is given by $v_r^q$, and the propagation path difference between the $r_m$-th activated port and $O_r$ is expressed as:
{
\setlength\abovedisplayskip{3pt}
\setlength\belowdisplayskip{3pt}
\begin{multline}\label{eq:dr}
d_r^q(r_m)=\\
\sqrt{{v_r^q}^2+(y_{{\rm U},m})^2-2 v_r^q y_{{\rm U},m}\cos\left({\frac{\pi}{2}-\theta_r^q}\right)}-v_r^q.
\end{multline}}

Consequently, we can simplify \eqref{eq:dr} using the Talor's approximation $\sqrt{1+x}\approxeq 1+\frac{1}{2}x-\frac{1}{8}x^2$ for small $x$ as:
\vspace{-2mm}
\begin{equation}
d_r^q(r_m)=-y_{{\rm U},m}\sin{\theta_r^q}-\frac{(y_{{\rm U},m})^2\sin^2{\theta_r^q}}{2v_r^q}.
\end{equation}

The signal phase difference between the $r_m$-th activated port and $O_r$ in the $q$-th receive path is obtained by $2\pi\rho_r^q(r_m)/\lambda$. Thus, the receive field response vector can be written as:
\begin{equation}
\boldsymbol{b}(r_m)\triangleq\left[e^{j\frac{2\pi}{\lambda}d_r^1(r_m)},\ldots, e^{j\frac{2\pi}{\lambda}d_r^{V_r}(r_m)}\right]^T\in \mathbb{C}^{V_r\times1}.
\end{equation}
The field response matrix of $m_a$ activated ports is given by:
\begin{equation}
\boldsymbol{B}(\boldsymbol{r})\triangleq[\boldsymbol{b}(r_1),  \ldots, \boldsymbol{b}(r_{m_a})]\in\mathbb{C}^{V_r\times m_a}.
\end{equation}

In the proposed system, we define the path response matrix from $O_t$ to $O_r$ as $\boldsymbol{O}\in\mathbb{C}^{V_r\times V_t}$, where $O_{q,p}$ refers to the response coefficient between the $p$-th transmit path and the $q$-th receive path. For each response coefficient, we assume that $O_{q,p} \overset{\mathrm{iid}}{\sim} \mathcal{N}(0,\alpha^2)$. Finally, the end-to-end channel matrix, $\boldsymbol{G}(\boldsymbol{r})$, is given: 
\vspace{-2mm}
\begin{equation}
\boldsymbol{G}(\boldsymbol{r})=\boldsymbol{B}^H(\boldsymbol{r})\boldsymbol{O}\boldsymbol{A}\in \mathbb{C}^{m_a\times N}.
\vspace{-1mm}
\end{equation}

By taking the expectation of the channel response matrix $\boldsymbol{O}$ over a period of time, we can obtain the achievable transmission rate as      
\vspace{-2mm}
\begin{equation}
R_0=\mathbb{E}_{\boldsymbol{O}}\left\{\log \det\left(\boldsymbol{\mathrm{I}}_{m_a}+\frac{1}{\sigma^2}\boldsymbol{G}(\boldsymbol{r})\boldsymbol{Q}\boldsymbol{G}^H(\boldsymbol{r})\right)\right\}.
\end{equation}
\vspace{-0.5em}
\subsection{Semantic Compression Model}\label{sec:B}
{ In the proposed system, the BS performs semantic extraction on the transmitted data set $\mathcal{X}$ via its local probability graph \cite{10333452}. Specifically, probability graph summarizes statistical insights from diverse knowledge graphs, expanding the conventional triplet knowledge graph by introducing the dimension of relational probability, while the BS statistically calculating occurrences of various relations between given entities, and filter those relations with high occurrence probability or conditional probability. In this way, the size of sent data is reduced.} The result of semantic extraction  $\mathcal{Y}$ will then be transmitted through the antenna arrays in the BS. To demonstrate the impact of semantic compression, semantic compression rate, defined as $\rho=\frac{\mathrm{size}(\mathcal{Y})}{\mathrm{size}(\mathcal{X})}$, is introduced, where $\mathrm{size}(\cdot)$ indicates a function measuring the size of the data. The equivalent rate of the entire system can be given as:
\vspace{-2mm}
\begin{equation}
    R=\frac{1}{\rho}R_0.
\end{equation}

Semantic extraction inevitably entails power consumption. {  According to \cite{10333452}, the computational load for the considered system is formulated as:
\vspace{-2mm}
\begin{equation}\label{eq:gn}
    c\left(\rho\right)=\left\{\begin{array}{l}
        A_1\rho +B_1, D_1\leq \rho \leq 1, \\
        A_2\rho +B_2, D_2\leq \rho < D_1, \\
        \vdots \\
        A_S\rho +B_S, D_S\leq \rho < D_{S-1},
    \end{array}\right.
    \vspace{-1mm}
\end{equation}
where $s \in \{1,2,\ldots,S\}$ is the index of each segment,  $B_s$ represents the intercept and $A_s$ denotes the slope of each segment, which  subjects to $0>A_1>A_2>\cdots>A_S$. $D_s$ serves as the partition of the computation load function. This piece-wise function is the consequence of complex conditional probability matrices required in each round of semantic extraction. As the quantity of information increases, a discernible pattern emerges and we can derive the parameters of \eqref{eq:gn}. Generally, this function reflects that a reduction in the compression rate $\rho$ increase the power consumption. }

With \eqref{eq:gn}, we can calculate the compression power consumption for the semantic extraction:
\vspace{-2mm}
\begin{equation}\label{eq:pc_}
    P_c = c(\rho)p_0,
    \vspace{-2mm}
\end{equation}
where $p_0$ stands for the positive computation power coefficient. 

By combining \eqref{eq:pc_} and the power consumption of the BS $\mathrm{tr}(\boldsymbol{Q})$, we can derive the power constraint:
\vspace{-1mm}
\begin{equation}
    P_c+\mathrm{tr}(\boldsymbol{Q})\leq P_{\max},
    \vspace{-1mm}
\end{equation}
where $P_{\max}$ stands for the total power that can be allocated to transmission and computation.
\vspace{-0.5em}
\subsection{Problem Formulation}\label{sec:D}
Our goal is to maximize the equivalent rate $R$ by optimizing $\boldsymbol{Q}$, $\rho$ and $\boldsymbol{r}$ jointly. Mathematically, the optimization problem can be formulated as
\vspace{-0.4mm}
{
\setlength\abovedisplayskip{3pt}
\setlength\belowdisplayskip{3pt}
\begin{subequations}
    \begin{align}
        \max_{\boldsymbol{Q},\boldsymbol{r},\rho} \quad& R= \mathbb{E}_{\boldsymbol{O}}\left\{
    \frac{1}{\rho}\log\det(\boldsymbol{I}_{m_a}+\frac{1}{\sigma^2}\boldsymbol{G}(\boldsymbol{r})\boldsymbol{Q}\boldsymbol{G}^{H}(\boldsymbol{r}))
    \right\}, \label{eq:maxr}\\
    \textrm{s.t.}\quad &\boldsymbol{r}=\left[r_1,\ldots,r_m,\ldots,r_{m_a}\right]^T\in\mathbb{Z}^{m_a\times1},\label{eq:cr1}\\
    &{ r_m\in\{1,2,\ldots,M\},m \in \{1,2,\ldots,m_a\}}\\
&r_1< r_2< \cdots < r_{m_a},\label{eq:cr3}\\
&\mathrm{tr}(\boldsymbol{Q})+P_c\leq P_{\rm max},\label{eq:proc1}\\
& P_c \geq 0,\label{eq:proc2}\\
&\boldsymbol{Q}\succeq \boldsymbol{0},\label{eq:proc3}
    \end{align}
\end{subequations}}
 where the constraints \eqref{eq:cr1} - \eqref{eq:cr3} ensure a legal port selection for the FAS, and the constraint \eqref{eq:proc1} guarantees the total power consumption does not exceed the power limit $P_{max}$. Besides, the constraints \eqref{eq:proc2} and \eqref{eq:proc3} incorporate the nature of computation power and beamforming matrix. In problem (16), the objective function is non-convex and involves the computation of expectation, which presents significant difficulty in solving the problem directly without extra simplification.
\section{Algorithm Design}
\subsection{Upper Bound for Equivalent Rate}
As demonstrated in objective function \eqref{eq:maxr}, the problem is inherently challenging to solve directly. To simplify \eqref{eq:maxr}, we first derive an upper bound to it using Jensen's inequality:
\vspace{-2mm}
{\setlength\abovedisplayskip{3pt}
\setlength\belowdisplayskip{3pt}\begin{align}
    R&\leq \overline{R} \nonumber\\
    &\triangleq \frac{1}{\rho}\log \det \left( \boldsymbol{\mathrm{I}}_{m_a}+\frac{1}{\sigma^2}\mathbb{E}_{\boldsymbol{O}}\left\{\boldsymbol{G}(\boldsymbol{r})\boldsymbol{Q}\boldsymbol{G}^H(\boldsymbol{r})\right\}\right)\nonumber\\
&=\frac{1}{\rho}\log\det \left(\boldsymbol{\mathrm{I}}_{m_a}+\frac{1}{\sigma^2}\boldsymbol{B}^H(\boldsymbol{r})\mathbb{E}_{\boldsymbol{O}}\left\{\boldsymbol{O}\boldsymbol{A}\boldsymbol{Q}\boldsymbol{A}^H\boldsymbol{O}^H\right\}\boldsymbol{B}(\boldsymbol{r})\right).\label{eq:ub}
\end{align}}

By employing the statistical feature of the channel response matrix $\boldsymbol{O}$, the expectation in \eqref{eq:ub} can be simplified as follows: 
{\setlength\abovedisplayskip{4pt}
\setlength\belowdisplayskip{3pt}\begin{equation}
    \mathbb{E}_{\boldsymbol{O}}\left\{\boldsymbol{O}\boldsymbol{A}\boldsymbol{Q}\boldsymbol{A}^H\boldsymbol{O}^H\right\}=\mathrm{tr}\left(\boldsymbol{A}\boldsymbol{Q}\boldsymbol{A}^H\right)\alpha^2\boldsymbol{\mathrm{I}}_{V_r}.
\end{equation}}

Moreover, the upper bound for the equivalent rate in \eqref{eq:ub} can be reduced to:
{\setlength\abovedisplayskip{4pt}
\setlength\belowdisplayskip{3pt}\begin{equation}
    \overline{R}=\frac{1}{\rho}\log\det \left(\boldsymbol{\mathrm{I}}_{m_a}+\frac{\alpha^2}{\sigma^2}\mathrm{tr}\left(\boldsymbol{A}\boldsymbol{Q}\boldsymbol{A}^H\right)\boldsymbol{B}^H(\boldsymbol{r})\boldsymbol{B}(\boldsymbol{r})\right).\label{eq:sr}
\end{equation}}

Though the calculation for expectation is avoided, the problem remains non-convex and hard to solve. Thus, an alternating optimization algorithm is proposed to address this issue.

\subsection{Optimization for Transmit Beamforming and Semantic Compression Ratio}
In this part of optimization, we assume the selection of the activated ports is fixed, and optimize the transmission covariance matrix $\boldsymbol{Q}$ and semantic compression rate $\rho$. Problem (16) is transformed into the following sub-problem
{\setlength\abovedisplayskip{3pt}
\setlength\belowdisplayskip{3pt}\begin{subequations}
    \begin{align}
        \max_{\boldsymbol{Q},\rho} \quad& \frac{1}{\rho}\log\det \left(\boldsymbol{\mathrm{I}}_{m_a}+\frac{\alpha^2}{\sigma^2}\mathrm{tr}\left(\boldsymbol{A}\boldsymbol{Q}\boldsymbol{A}^H\right)\boldsymbol{B}^H(\boldsymbol{r})\boldsymbol{B}(\boldsymbol{r})\right), \label{eq:sp1}\\
    \textrm{s.t.}\quad 
&(\ref{eq:proc1})-(\ref{eq:proc3}) \nonumber
    \end{align}
\end{subequations}}

It is evident that when the inequality constraint \eqref{eq:proc1} holds with equality, the equivalent rate is maximized with a certain $\boldsymbol{Q}$. So \eqref{eq:proc1} can be rewritten as:
{\setlength\abovedisplayskip{4pt}
\setlength\belowdisplayskip{3pt}\begin{equation}
    P_c=P_{\rm max}-\mathrm{tr}(\boldsymbol{Q}) \geq 0.\label{eq:pc}
    \vspace{-1mm}
\end{equation}}

We can derive the relation between $\boldsymbol{Q}$ and $\rho$ with \eqref{eq:pc}. From \eqref{eq:gn}, we rewrite the compression rate with $P_c$:
{\setlength\abovedisplayskip{4pt}
\setlength\belowdisplayskip{3pt}\begin{equation}
    \rho=\sum_{s=1}^{S}\frac{P_c/p_0-B_s}{A_s} \theta_{s},\label{eq:f}    
\end{equation}}
where $\sum_{s=1}^{S} \theta_s=1, \theta_s \in \{0,1\}, \forall s \in \{1,2,\ldots,S\}$. $\theta_s$ indicates the segment of computation load function corresponding to the computation power $P_c$.

Combining \eqref{eq:pc}, problem (20) can be simplified to:
{ \setlength\abovedisplayskip{3pt}
\setlength\belowdisplayskip{3pt}
\begin{subequations}
    \begin{align}
        \max_{\boldsymbol{Q},\boldsymbol{\theta}} \quad& \frac{\log\det \left(\boldsymbol{\mathrm{I}}_{m_a}+\frac{\alpha^2}{\sigma^2}\mathrm{tr}\left(\boldsymbol{A}\boldsymbol{Q}\boldsymbol{A}^H\right)\boldsymbol{B}^H(\boldsymbol{r})\boldsymbol{B}(\boldsymbol{r})\right)}{\sum_{s=1}^{S}\frac{(P_{\rm max}-\mathrm{tr}(\boldsymbol{Q}))/p_0-B_s}{A_s} \theta_{s}} ,\label{eq:sp1_}\\
        \textrm{s.t.}\quad 
&\mathrm{tr}(\boldsymbol{Q})\leq P_{\rm max},\label{eq:sp1fc1}\\
\vspace{-0.5mm}
&  P_{\rm max}-\sum_{s=1}^{S} \theta_s (A_s D_{s}+B_s) \nonumber \\\vspace{-0.5mm}
& \leq \mathrm{tr}(\boldsymbol{Q})\leq  P_{\rm max}-\sum_{s=1}^{S} \theta_s(A_s D_{s-1}+B_s) ,\label{eq:segc}\\
\vspace{-0.5mm}
&\boldsymbol{Q}\succeq \boldsymbol{0} ,\\
\vspace{-0.5mm}
&\sum_{s=1}^{S} \theta_s=1 ,\\
\vspace{-0.5mm}
&\theta_s \in \{0,1\}, \forall s \in \{1,2,\ldots,S\},\label{eq:sp1fc6}
    \end{align}
\end{subequations} }

where the constraint \eqref{eq:segc} is derived from \eqref{eq:proc1}, and $D_0=1$ represents compression rate $\rho=1$.

To transform problem (23) into a linear combination, a balanced factor $\tau$ is introduced. We denote the numerator of \eqref{eq:sp1_} as $f(\boldsymbol{Q})$ and the denominator as $g(\boldsymbol{Q})$, and then rewrite the optimization problem (23) as follows:
\vspace{-1mm}
\begin{subequations}
    \begin{align}
        \max_{\boldsymbol{Q}} \quad& f(\boldsymbol{Q})-\tau g(\boldsymbol{Q}),\label{eq:sp1f}\\
        \textrm{s.t.}\quad 
        &(\ref{eq:sp1fc1})-(\ref{eq:sp1fc6}).\nonumber
        \vspace{-1mm}
    \end{align}
\end{subequations}

We then develop an iterative algorithm to solve problem (24). First, we set an initial value $\boldsymbol{Q}^{(0)}$, which satisfies constraints (\ref{eq:sp1fc1})-(\ref{eq:sp1fc6}). Then, we keeps updating $\tau$ by letting $\tau^{(i)}=\frac{f(\boldsymbol{Q}^{(i)})}{g(\boldsymbol{Q}^{(i)})}$, where $i$ is the iteration index. With $\tau^{(i)}$ known, problem (26) becomes a convex problem and can be efficiently solved by existing optimization tools such as CVX. After solving problem (24), $\boldsymbol{Q}^{(i+1)}$ can be obtained. The iteration process will stop when the value of \eqref{eq:sp1f} falls within $[-\epsilon_1, \epsilon_1]$, where $\epsilon_1$ is a preset threshold. The details of the algorithm is shown in Algorithm~\ref{al:1}.
\vspace{-0.5em}
\begin{algorithm}
    \caption{Optimization for transmit covariance matrix and compression rate}
    \begin{algorithmic}\label{al:1}
    \STATE \textbf{Initialize:} $\boldsymbol{Q}^{(0)}$, threshold $\epsilon_1$, iteration index $i=0$, $\boldsymbol{\theta}^{(0)}$.
    \REPEAT
        \STATE Set iteration index $i=0$.
        \REPEAT
            \STATE $\tau^{(i)}=\frac{f(\boldsymbol{Q}^{(i)})}{g(\boldsymbol{Q}^{(i)})}$
            \STATE Calculate $\boldsymbol{Q}^{(i+1)}= \arg \max_{\boldsymbol{Q}} \;f(\boldsymbol{Q}^{(i)})-\tau^{(i)}g(\boldsymbol{Q}^{(i)})$
            \STATE Set $i=i+1$
        \UNTIL{$\lvert f(\boldsymbol{Q}^{(i)})-\tau^{(i-1)}g(\boldsymbol{Q}^{(i)}) \rvert \leq \epsilon_1$}
        \STATE Obtain $\boldsymbol{\theta}^{(i)}$
    \UNTIL{All situations of $\boldsymbol{\theta}$ are considered.}
        \STATE \textbf{Output:} 
        Transmit covariance matrix $\boldsymbol{Q}$, compression rate $\rho$
    \end{algorithmic}
\end{algorithm}
\vspace{-0.5em}

It is notable that the result of Algorithm 1 may not be the optimal solution to problem (23), but it is still an acceptable sub-optimal solution owing to good convergence and appropriate complexity.
\vspace{-0.5em}
\subsection{Optimization for Port Selection}

With transmit covariance matrix $\boldsymbol{Q}$ and compression rate $\rho$ known, problem (16) can be re-formulated as:
\vspace{-2mm}
\begin{subequations}
    \begin{align}
        \max_{\boldsymbol{r}} \quad &\frac{1}{\rho}\log\det \left(\boldsymbol{\mathrm{I}}_{m_a}+\frac{\alpha^2}{\sigma^2}\mathrm{tr}\left(\boldsymbol{A}\boldsymbol{Q}\boldsymbol{A}^H\right)\boldsymbol{B}^H(\boldsymbol{r})\boldsymbol{B}(\boldsymbol{r})\right), \label{eq:r}\\
        \textrm{s.t.}\quad &(\ref{eq:cr1})-(\ref{eq:cr3}).\nonumber
    \end{align}
\end{subequations}

By introducing $\gamma=\frac{\alpha^2}{\sigma^2}\mathrm{tr}\left(\boldsymbol{A}\boldsymbol{Q}\boldsymbol{A}^H\right)$, \eqref{eq:r} can be simplified to:
\vspace{-2mm}
\begin{subequations}
    \begin{align}
        \overline{R}&=\frac{1}{\rho}\log\det \left(\boldsymbol{\mathrm{I}}_{m_a}+\gamma\boldsymbol{B}^H(\boldsymbol{r})\boldsymbol{B}(\boldsymbol{r})\right) \nonumber\\
        &=\frac{1}{\rho}\log\det \left(\boldsymbol{\mathrm{I}}_{V_r}+\gamma\sum_{m=1}^{m_a}\boldsymbol{b}(r_m)\boldsymbol{b}^H(r_m)\right)\label{eq:bm}.
    \end{align}
\end{subequations}

It is difficult to solve the integer optimization problem (25). We seek to optimize a single element $r_m$ in $\boldsymbol{r}$ instead. With this motivation, we remove the $m$-th column from matrix $\boldsymbol{B}$, and reconstruct the remaining matrix:
\vspace{-1mm}
\begin{equation}
    \overline{\boldsymbol{B}}_m=[\boldsymbol{b}(r_1),\ldots,\boldsymbol{b}(r_{m-1}),\boldsymbol{b}(r_{m+1}),\ldots,\boldsymbol{b}(r_{m_a})]. \label{eq:bmb}
\end{equation}

Rewrite \eqref{eq:bm} with \eqref{eq:bmb}, we can separate the terms with $r_m$:
\vspace{-1mm}
\begin{subequations}
    \begin{align}
        \overline{R}&=\log \det\left(1+\gamma\boldsymbol{b}^H(r_m)\left(\boldsymbol{I}_{V_r}+\gamma \overline{\boldsymbol{B}}_m\overline{\boldsymbol{B}}^H_m\right)^{-1}\boldsymbol{b}(r_m)\right) \nonumber\\
        &+\log \det\left(\boldsymbol{I}_{V_r}+\gamma \overline{\boldsymbol{B}}_m\overline{\boldsymbol{B}}^H_m\right).\label{eq:rbm}
    \end{align}
\end{subequations}

Now, we consider that $\overline{\boldsymbol{B}}_m$ is determined, and optimize \eqref{eq:r}. The problem becomes:
\vspace{-2mm}
\begin{subequations}
    \begin{align}
        \max_{r_m} \quad &\boldsymbol{b}^H(r_m)\left(\boldsymbol{I}_{V_r}+\gamma \overline{\boldsymbol{B}}_m\overline{\boldsymbol{B}}^H_m\right)^{-1}\boldsymbol{b}(r_m), \label{eq:rm}\\
        \textrm{s.t.}\quad &(\ref{eq:cr1})-(\ref{eq:cr3}).\nonumber
    \end{align}
\end{subequations}

If the total number of ports $M$ and the number of activated ports $m_a$ are not excessive (for example, both less than one hundred), it is possible to go through all valid values of $r_m$ to find the most suitable one. The details of the algorithm are displayed in Algorithm~\ref{al:2}.
\vspace{-0.5em}
\begin{algorithm}
    \caption{Alternate optimization for port selection, transmit covariance matrix and
compression rate}
    \begin{algorithmic}\label{al:2}
        \STATE \textbf{Initialize:} $\boldsymbol{Q}^{(0)}$, iteration index $i=0$, threshold $\epsilon_2$, $\boldsymbol{r}^{(0)}$, value of objective function $\eta^{(0)}(\boldsymbol{Q}, \boldsymbol{r})$.
        \REPEAT
        \STATE Obtain $\boldsymbol{Q}$ from solving \eqref{eq:sp1f}
        \FOR{$m$=1 to $m_a$}
            \STATE Obtain $\overline{\boldsymbol{B}}_m$ from \eqref{eq:bmb}
            \STATE Find the $r_m$ which maximizes \eqref{eq:rm}
        \ENDFOR
        \STATE Set $i=i+1$
        \STATE Calculate $\eta^{(i)}(\boldsymbol{Q}, \boldsymbol{r})$
        \UNTIL{$|\eta^{(i)}(\boldsymbol{Q}, \boldsymbol{r})-\eta^{(i-1)}(\boldsymbol{Q}, \boldsymbol{r})|\leq \epsilon_2$}
    \end{algorithmic}
\end{algorithm}

{  Regarding the convergence of Algorithm.2, subproblem (23) is solved using Dinkelbach's transform, which is proved to converge with given accuracy when the objective fraction is a concave-convex ratio \cite{10.1287/mnsc.13.7.492}, while subproblem (25) is solved using an enumerate-like method. The value of (21) increases monotonously during the optimization and as (21) is boundary, the overall algorithm is bound to converge.}

The semi-define programming problem (23) can be solved with a worst-case time complexity $O(N^{4.5}\log(1/\epsilon_1))$. After applying Dinkelbach's transform, the time complexity for optimizing transmit covariance matrix and semantic compression ratio arrives $O(\frac{N^{4.5}\log(1/\epsilon_1)}{\epsilon_2})$. In the worst case, we need to go over all combinations to find the best port selection. With consider that $m_a$ is usually much smaller than $M$, the worst time-complexity for port selection is $O(M^{m_a})$. So the worst-case time-complexity for the Algorithm 2 is $O(\frac{N^{4.5}\log(1/\epsilon_1)}{\epsilon_2}M^{m_a})$.

\section{Simulation Results}
This section presents the results of simulations conducted on the proposed algorithm. Three additional schemes have been conducted as comparison, as described below:
\begin{list}{\labelitemi}{\leftmargin=1em}
    \setlength{\topmargin}{0pt}
    \setlength{\itemsep}{0em}
    \item \textbf{Random-FAS-semantic:} The activated ports in the FAS are  selected randomly. Semantic extraction is still employed. Algorithm \ref{al:1} is used to optimize semantic compression rate and transmit covariance matrix.
    \item \textbf{FAS-non-semantic:} A typical FAS system without semantic extraction. Algorithm from \cite{Chen2024JointBA} is adopted to optimize the transmit covariance matrix and port selection in FAS.
    \item \textbf{Conventional:}  A conventional receive antenna system at the user end, with two static antennas whose coordinates are $(0,\frac{1-M}{2})$ and $(0,\frac{2M-1}{2})$, respectively. No semantic extraction will be performed. 
\end{list}

In simulation, the number of transmit antennas $N$ is set to 20, the number of ports in FAS $M$ is 35, and the number of activated ports $m_a$ is $5$. The carrier wavelength $\lambda$ is set to 4mm, and the spacing between transmit antennas $d_{BS}$ and ports in FAS $d_{U}$ are both $\lambda/2$. Number of transmit paths $V_t$ and number of receive paths $V_r$ are both 3. The average noise power is set to 3dBm, and the average variance of response coefficient is $1/V_r$. The solution accuracy $\epsilon_1$ and $\epsilon_2$ is both $10^{-5}$.
 For AoDs $\theta_t^p, p\in[1,V_t]$ and AoAs $\theta_r^q, q\in[1,V_r]$, they are both distributed uniformly and randomly among $[-\frac{\pi}{2},\frac{\pi}{2}]$. The signal-to-noise rate (SNR) is defined as $P_{\max}/\sigma^2$. { Parameters for semantic compression are set according to \eqref{eq:gn}, and the power coefficient $p_0$ is set to 1.}

\begin{figure}
    \centering
    \includegraphics[width=0.65\linewidth]{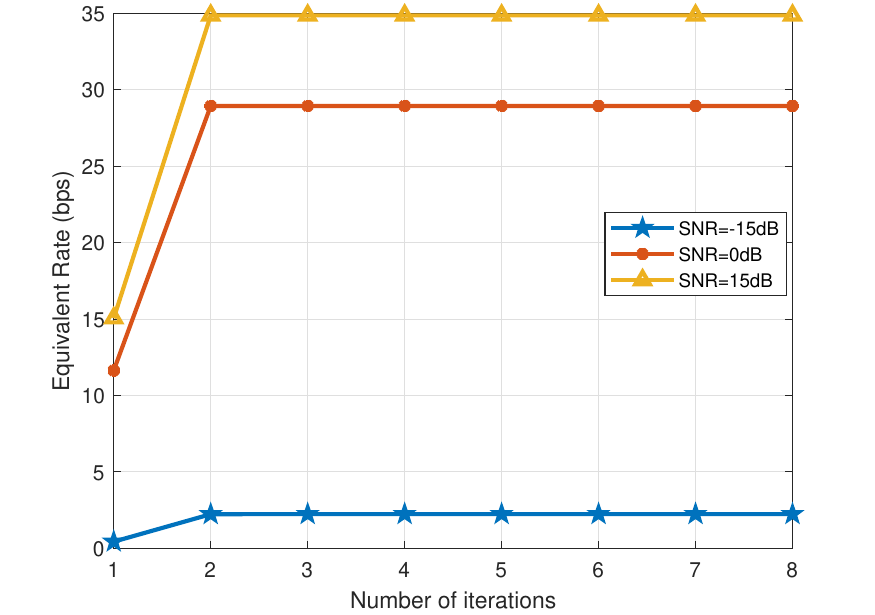}
     \vspace{-0.5em}
    \caption{Number of iterations versus equivalent rate.}
   \vspace{-1em}
    \label{fig:1}
\end{figure}

Fig.~\ref{fig:1} plots the convergence of the proposed algorithm with varying SNRs.  It is evident that the proposed algorithm reaches convergence fast, and the equivalent rate after initial optimization round is notably close to the final result. This demonstrates the time-efficiency of the proposed algorithm.

\begin{figure}
    \centering
    \includegraphics[width=0.65\linewidth]{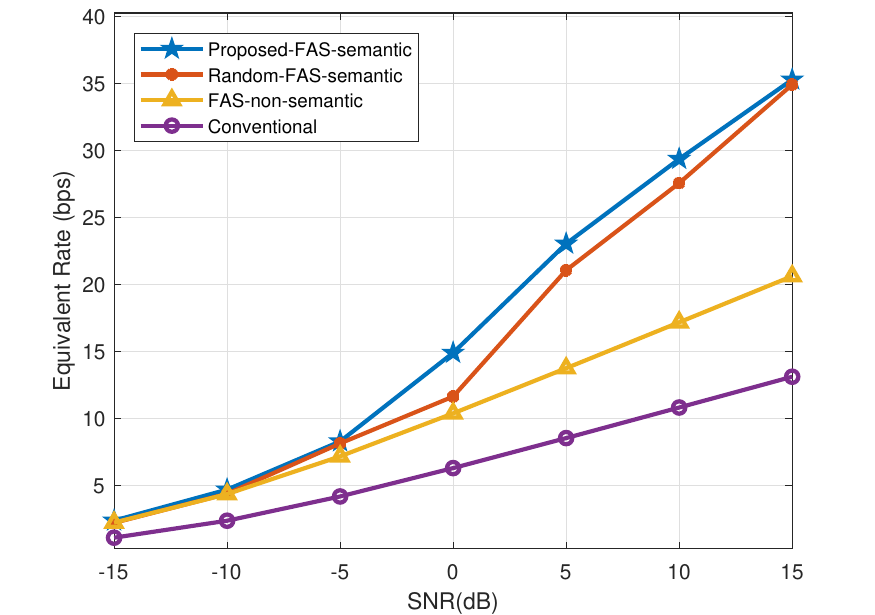}
    \vspace{-0.5em}
    \caption{SNR versus equivalent rate.}
    \label{fig:2}
    \vspace{-1em}
\end{figure}

Fig.~\ref{fig:2} demonstrates the performance of the four schemes with varying SNRs. As SNR increasing, all schemes show a consistent growth in performance. {  Notably, the proposed scheme and scheme `Random-FAS-semantic' demonstrate a more pronounced rise, increasing 137\% and 205\% respectively, when SNR rises from 0dB to 15dB. In contrast, scheme `FAS-non-semantic' only  exhibits a mere 98\% increase and scheme `Conventional' rises 108\%. When SNR=15dB, the proposed scheme exhibits a 71\% enhancement in performance relative to scheme `FAS-non-semantic', which is caused by the amplification effect of semantic compression ratio when the channel is preferable and less transmit power is needed. The same advantage appears on scheme `Random-FAS-semantic' as well. However, when SNR=0dB, scheme `Random-FAS-semantic' only outperforms scheme `FAS-non-semantic' by 12\%, while this ratio being 43\% for the proposed scheme. In the absence of rational port selection policy, scheme `Random-FAS-semantic' exhibits persistent underperformance in comparison to the proposed scheme, with the performance gap fluctuating randomly.}
\vspace{-0.5em}
\section{Conclusion}
In this paper, we have proposed a semantic-communication-assisted near-field FAS system, and then presented an alternating algorithm to jointly optimize the BS beamforming, port selection in FAS and semantic compression rate, aiming at maximizing the equivalent transmission rate. Numerical results have been provided to prove the effectiveness of the proposed algorithm.
\vspace{-1em}
\bibliographystyle{IEEEtran}
\footnotesize
\bibliography{ref}
\end{document}